\documentclass[prx,aps,twocolumn,showpacs,preprintnumbers,amsmath,amssymb]{revtex4-1}

\usepackage{graphicx}
\usepackage{dcolumn}

\usepackage[usenames,dvipsnames]{color}

\def\EF{$E_\textrm{F}$}
\def\SVO{SrVO$_3$}
\def\SNO{SrNbO$_3$}
\def\SVNO{Sr$_2$VNbO$_6$}
\def\lSVNO{$l$--Sr$_2$VNbO$_6$}
\def\rSVNO{$r$--Sr$_2$VNbO$_6$}

\def\t2g{t$_{\textrm{2g}}$}

\begin{document}

\title{Cation Order Control of Correlations in Double Perovskite Sr$_2$VNbO$_6$} 
\author{Arpita \surname{Paul}}
\author{Turan \surname{Birol}}
\email{tbirol@umn.edu}
\affiliation{Department of Chemical Engineering and Materials Science, University of Minnesota, Minneapolis, Minnesota 55455, USA}
\date{\today}

\begin{abstract}
	Double perovskites extend the design space for new materials, and they often host phenomena that don't exist in their parent perovskite compounds. 
	Here, we present a detailed first principles study of the correlated double perovskite Sr$_2$VNbO$_6$, where inter-cationic charge transfer and strength of electronic correlations depend strongly on the cation order. 
	By using Density Functional Theory + Embedded Dynamical Mean Field Theory, we show that this compound has a completely different electronic structure than either of its parent compounds despite V and Nb being from the same group in the periodic table. 
	We explain how the electronic correlations' effect on the crystal structural parameters determines on which side of the Hund's metal--Mott insulator transition the material is. 
	Our results demonstrate the emergence of Hund's metallic behavior in a double perovskite that has $d^1$ parents, and underlines the importance of electronic correlation effects on the crystal structure. 
\end{abstract}
\maketitle

\section{Introduction}

Transition metal oxides (TMOs) are in the focus of great interest as they host a wide variety of electronic phenomena ranging from metal-insulator transitions, charge/orbital/spin ordering, multiferroicity, colossal magnetoresistance, different types of magnetism, and high temperature superconductivity \cite{Tokura2000,Cheong2007,Rao1989,Khomskii1997}. Many of these emergent properties and rich phase diagrams arise from the interplay between charge, spin, orbital and lattice degrees of freedom. Strong Coulomb interactions between electrons in the $d$ orbitals of transition metals often lead to strong electronic correlations, and the directional nature of these orbitals leads to strong coupling between the electronic wave function and the lattice degrees of freedom \cite{Khomskii2014_book}. This makes it possible for seemingly very similar compounds to have very different properties that can be tuned easily via external fields, or chemical substitution. 

\begin{figure}
\centering
\includegraphics[width=1.0\linewidth]{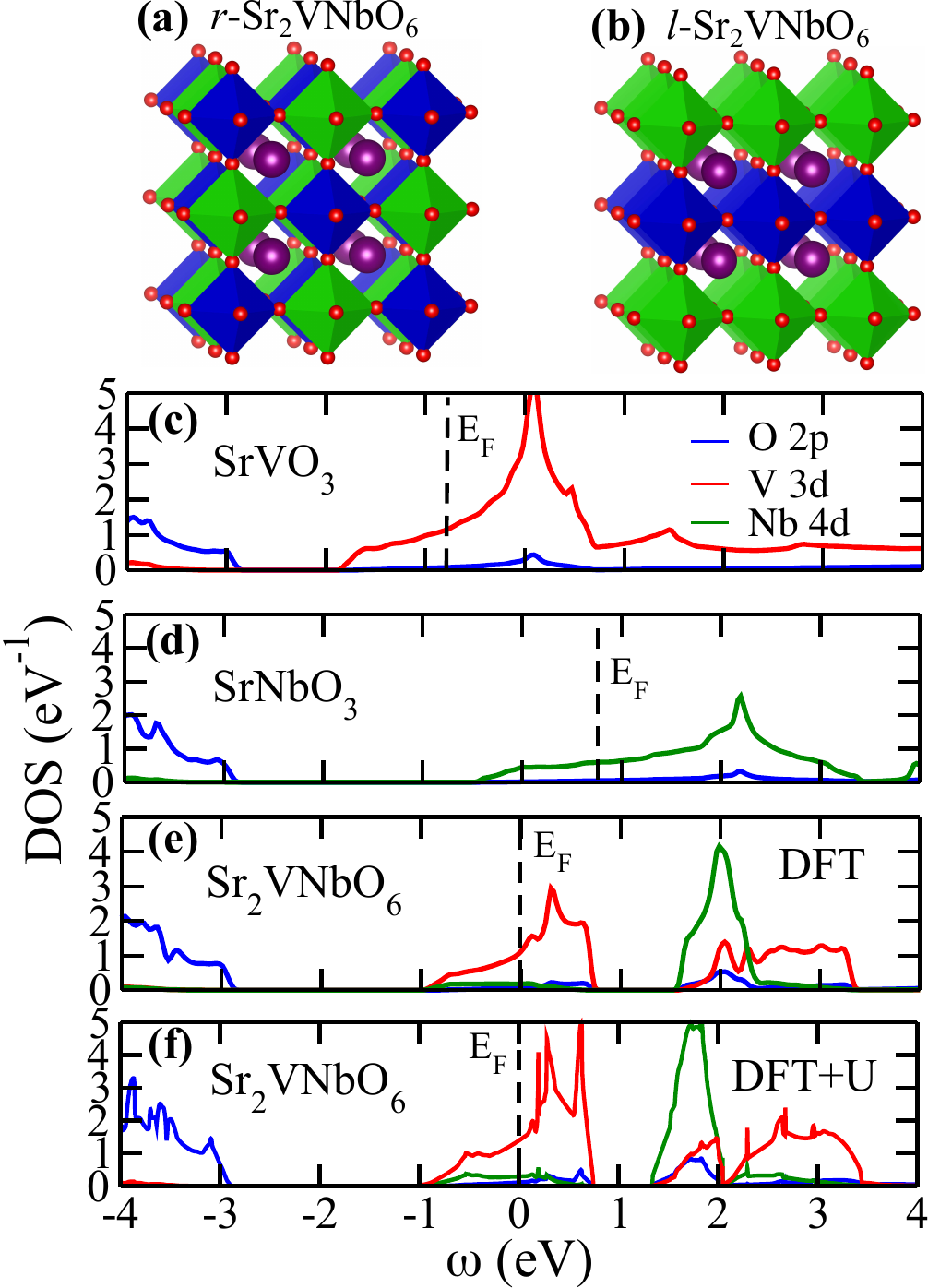}
	\caption{(a) Rocksalt and (b) layered ordered double perovskite Sr$_2$VNbO$_6$. VO$_6$ and NbO$_6$ octahedra are presented in blue (dark) and green (light) colors. (c) Density of states of SrVO$_3$, (d) SrNbO$_3$, and (e) rocksalt ordered Sr$_2$VNbO$_6$ calculated using DFT. The O 2p levels in different compounds are aligned with each other. (f) Non-spin-polarized DFT+U calculations on the rocksalt ordered double perovskite give qualitatively similar densities of states to calculations with no U. }
\label{fig:str_dos} 
\end{figure}

Perhaps the best demonstration of the richness of TMOs is provided by the perovskite structure: almost any phenomenon observed in the solid state is realized in at least one ABO$_3$ perovskite oxide. 
%
%
The phase diagrams of perovskites can be further expanded by considering \textit{double perovskites}, where one of the cations is partially substituted in an ordered fashion \cite{Vasala2015, King2010}. For example, in the most common form of B-site ordered  A$_2$BB$'$O$_6$ double perovskites, there are 2 transition metals on the B-site that are alternating at every other unit cell (Fig. \ref{fig:str_dos}a). Multiple B site transition metal cations provide one more degree of freedom to realize different electronic phases
such as the rare combination of ferromagnetism with insulating behavior that is rather commonplace in double perovskites like La$_2$MnNiO$_6$ \cite{Matar2007}. Other phenomena such as multiferroicity, frustrated antiferromagnetism, magneto-optic properties, and spin liquid related phenomena have been observed in double perovskites with parent compounds that don't display the same properties \cite{Kobayashi1998,Kato2002,Sakai2007,Das2008,Mustonen2018}. 

The ordering of different metal ions in double perovskites greatly affects electronic and structural properties \cite{King2010}. For example the degree of cation order often determines the strength of relaxor characteristics that ferroelectric double perovskites display \cite{Davies2008}, and the ferromagnetic properties of the half-metal  Sr$_2$FeMoO$_6$ are affected significantly from cation disorder \cite{Meneghini2009, Sarma2001}. It is also possible to have different types of cation orders, where the B and B$'$ ions arrange themselves in a rocksalt, layered, or columnar fashion \cite{King2010}, and a particular stoichiometry can in principle give rise to very different properties depending on the cation order because of the different connectivity of BO$_6$ octahedra. (The rocksalt ordering (Fig. \ref{fig:str_dos}a) gives rise to BO$_6$ octahedra that are isolated from each other, whereas the layered ordering (Fig. \ref{fig:str_dos}b) gives rise to extended planes of connected BO$_6$ octahedra.) 

In this first principles study, we focus on the seemingly simple double perovskite \SVNO. Both parent compounds of \SVNO ~(\SVO ~and \SNO) ~are metallic perovskites with a single electron in partially filled $d$ shells of their B-site cations (V$^{4+}$ and Nb$^{4+}$) \cite{Paul2019_PRM}. Using Density Functional Theory + Embedded Dynamical Mean Field Theory (DFT + DMFT) \cite{Haule2010, Paul2019_review, Kotliar2004}, we show that this double perovskite has an electronic structure that is strongly intertwined with the cation order and crystal structural parameters (bond length disproportionation) in an exceedingly sensitive fashion. Even though V and Nb come from the same group in the periodic table, the difference between their electronegativites leads to an almost complete transfer of an electron from Nb to V in \SVNO, which results in formal valences closer to Nb$^{5+}$ and V$^{3+}$ instead of Nb$^{4+}$ and V$^{4+}$. The degree of this inter-cationic electron transfer depends on the type of the cation order (layered vs. rocksalt) present in the system. Depending on this cation order and the resulting changes in the bandwidths, the double perovskite \SVNO ~behaves either as a strongly correlated metal, which has part of its electronic correlations induced by the on-site Hund's coupling $J$ \cite{Deng2019, Haule2009}, or a Mott insulator, despite the fact that both parent compound \SVO ~and \SNO ~are mildly correlated Fermi liquids. Our results also underline the effect of the paramagnetic moments on the crystal structure and show that in order to get the Mott insulating phase the crystal structure parameters need to be calculated using DFT+DMFT, rather than DFT or DFT+$U$ as is commonly done due to the computational cost of structural predictions from DFT+DMFT.

\section{Methods} 

Determination of lattice constants and internal atomic coordinates at the DFT+U level is performed using using density functional theory and projector augmented wave formalism as implemented in the Vienna Ab Initio Simulation Package (VASP) unless stated otherwise \cite{VASP1, VASP2}. DFT+$U$ correction was employed for both transition metals with $U=4$~eV for V and $U=1$~eV for Nb. These values of $U$ obtained by calculating the bulk lattice constants of \SVO ~and \SNO ~that match the experimentally reported values \cite{Rey1990, Ridgley1955, Dudarev1998}. The DFT bandstructures reported are calculated using the linearized augmented plane wave approach as implemented in the Wien2K code \cite{WIEN2K}, and no DFT+$U$ correction. The DFT+DMFT calculations are performed using the eDMFT software package \cite{Haule2010, Haule2016Forces}. All DMFT calculating are performed at 290~K. Both V and Nb atoms are treated as impurities with on-site Hubbard $U$ values of $U=10$~eV for V and $U=6$~eV for Nb. These values are significantly higher than the ones used for DFT+U because of the different screening processes taken explicitly into account in DFT+DMFT calculations. We note that the suitable $U$ values are also implementation dependent: Our values are larger than typical values used for different DFT+DMFT implementations that use a Wannier based approach, rather than a local projector based one. These larger values are explicitly and extensively tested for various, especially 3d, transition metals, and are known to capture quantities such as bandwidth renormalization correctly \cite{Haule2014, Paul2019_PRM,Haule2015_energies}. The on-site Hund's $J$ value, on the other hand, is not strongly screened and is expected to be very similar in all of these methods. We use $J=0.7$~eV, except in cases where we repeat calculations with different values of $J$ to see how the strength of correlations depend on it. All of the reported DFT, DFT+U, and DFT+DMFT calculations preserve time reversal symmetry, and hence correspond to paramagnetic or diamagnetic phases, which are expected at room temperature. Further details of computational methods are provided in the supplementary material \cite{Supplement}. 

\section{Results}

\subsection{Rocksalt ordered double perovskite} 
Both \SVO ~and \SNO ~have the cubic perovskite structure with no structural distortions such as octahedral rotations at room temperature \cite{Rey1990, Ridgley1955}. 
%
%
Their \t2g bands are well separated from the oxygen p bands by $\sim$1 eV in \SVO ~and $\sim$2.5 eV in \SNO, as shown in Fig. \ref{fig:str_dos}c-d. This separation is determined by a combination of the bandwidths, crystal field splittings, and the electronegativities of V and Nb cations: V$^{4+}$ is more electronegative than Nb$^{4+}$ \cite{Li2006}, and hence d bands of V are lower in energy than those of Nb.
This large difference between the energies of the \t2g bands in the parent compounds lead to an interesting electronic structure in the rocksalt ordered double perovskite \SVNO. In Fig. \ref{fig:str_dos}e, we show the density of states of the rocksalt ordered \SVNO, henceforth referred to as \rSVNO, at the DFT level calculated using the DFT+$U$-optimized crystal structure. 
Only the vanadium \t2g bands cross the Fermi level, and niobium d shell is formally empty.
An electron is transferred from Nb to V compared to the parent perovskites, in other words, instead of the 4+ valences of transition metals in SrV$^{4+}$O$_3$ and SrNb$^{4+}$O$_3$, the double perovskite has Sr$_2$V$^{3+}$Nb$^{5+}$O$_6$. 
(Use of DFT+U without breaking time reversal symmetry does not change the main features of the DOS, as shown in Fig. \ref{fig:str_dos}f.)

Transfer of charge between B site cations in A$_2$BB'O$_6$ double perovskites is not uncommon \cite{Vasala2015, Chen2017b}. For example, Mo is in a higher valence state in Sr$_2$FeMoO$_6$ and Sr$_2$VMoO$_6$ than in SrMoO$_3$ \cite{Kuepper2008, Karen2006}. In Ba$_2$VFeO$_6$, Mott multiferroicity is predicted to emerge as a result of the charge transfer from V to Fe \cite{Chen2017a}. 
Short period (LaTiO$_3$)$_1$/(LaNiO$_3$)$_1$ heterostructures, which can be considered as layered double perovskites, also display similar charge transfer \cite{Chen2013}. 
While, unlike these examples, V and Nb are from the same group, a significant difference in the charge states of V and Nb is experimentally observed in SrV$_{1-x}$Nb$_x$O$_3$ solid solutions as well \cite{Miruszewski2018}; but any implications thereof in ordered \SVNO ~double perovskites are not addressed in the literature yet. 
%

\begin{figure}
\centering
\includegraphics[width=1.0\linewidth]{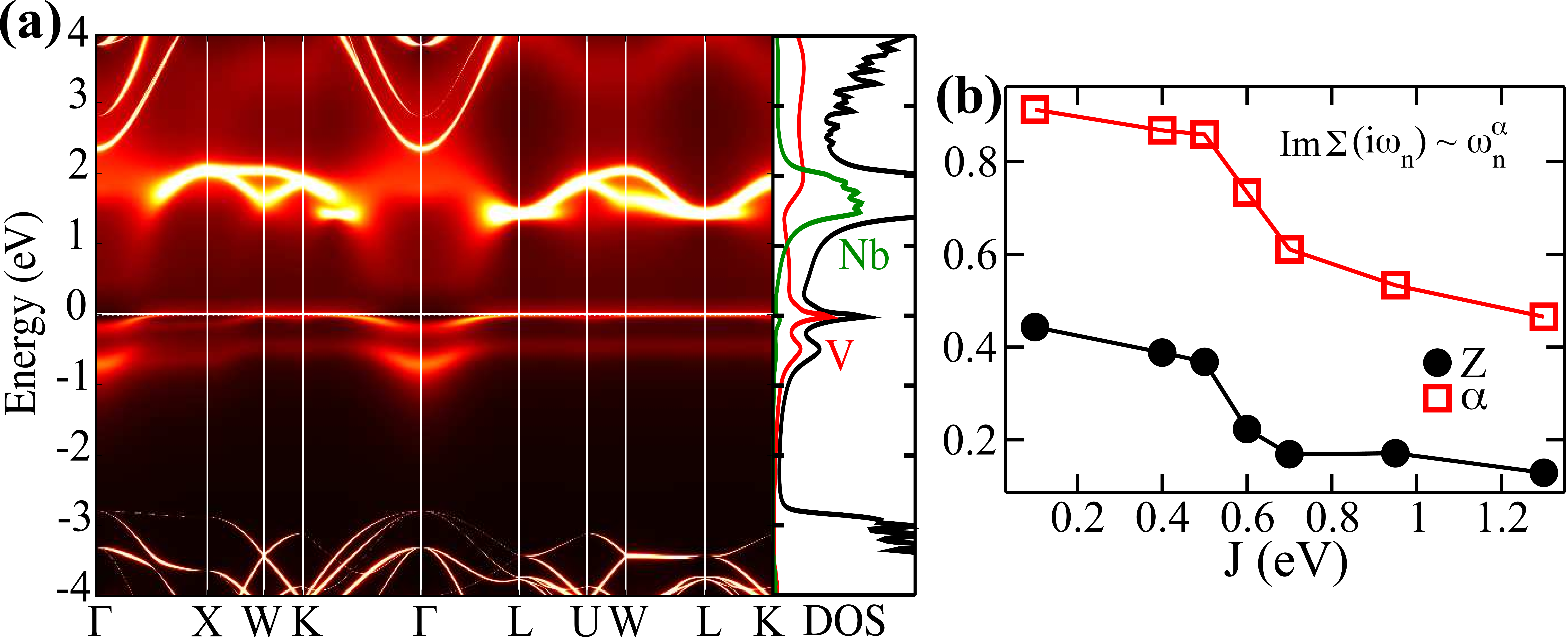}
		\caption{(a) The spectral function $S(k, \omega)$ and density of states of rocksalt ordered \rSVNO ~at room temperature calculated using DFT+DMFT. (b) Quasiparticle weight $Z$ and the exponent of the imaginary part of the self energy on the imaginary axis $\alpha$ of \rSVNO ~at room temperature for varying degrees of on-site Hund's coupling $J$. Hund's coupling in the range of $\sim0.1 -0.9$~eV is expected to describe the V ion well. 
	}
	\label{fig:rs_dmft}%
\end{figure}

In order to predict the electronic structure of \SVNO ~at room temperature, we performed DFT+embedded dynamical mean field theory (DFT+eDMFT) calculations \cite{Haule2010, Paul2019_review}. 
A simple measure of electronic correlation strength, especially for correlated metals that are close to a Fermi liquid phase, is the mass renormalization factor $Z$, which gives the ratio of the effective mass $m^*$ of the electrons to the band mass $m_b$ calculated from DFT, and is calculated from the slope of the DMFT electronic self energy near the Fermi level as 
\begin{equation}
	Z=\frac{m_b}{m^*}=\left.\left(1-\frac{\partial\textrm{Re} \Sigma(\omega)}{\partial \omega}\right)^{-1}\right|_{\omega=0}.
\end{equation}
Earlier dynamical mean field theory (DMFT) calculations and experiments agree that the parent compounds have mass renormalization factors of $Z_{\textrm{SrVO3}}\sim 0.5$ and $Z_{\textrm{SrNbO3}}\sim 0.7$ \cite{Paul2019_PRM, SNO2,Yoshida2005,Taranto2013}. (This value is very well established for SrVO$_3$, but it is harder to verify experimentally in SrNbO$_3$ due to possible Sr deficiency of the samples.) 
In Fig. \ref{fig:rs_dmft}a we present the spectral function and density of states of \SVNO. As expected, the V-\t2g bands near \EF ~are narrower compared to those in DFT, and hence the double perovskite (with the crystal structure from DFT+U) is a highly correlated metal with $Z=0.16$ at room temperature according to DFT+DMFT. It also behaves as a non-Fermi liquid: The exponent $\alpha$ of the self energy, which is $\alpha=1$ in ideal Fermi liquids, is $\alpha\sim 0.61$, which is a signature of Hund's metallic behavior \cite{Yin2012, Deng2019}.   

In order to elucidate the origin of strong correlations in double perovskite \SVNO, we repeat our DFT+eDMFT calculations using the same value of Hubbard $U$, but varying the value of Hund's coupling $J$. (Our results so far used $J=0.7$~eV, a typical value for V in our implementation.) The results, presented in Fig. \ref{fig:rs_dmft}b show that both $Z$ and $\alpha$ strongly depend on $J$. The value of $Z$ increases by more than 2-fold with decreasing $J$. There is also a large increase in $\alpha$ with reduced $J$, and its gets larger than $\sim0.90$ for small values of $J$. 
This behavior of $Z$ at room temperature is comparable to that observed in compounds such as SrRuO$_3$ and CaRuO$_3$, where the orbital degeneracy and the on-site Hund's coupling are important in determining the correlation strength \cite{Dang2015}. Thus, the double perovskite \SVNO ~can be classified as a Hund's metal -- but only if the DFT+$U$ determination of crystal structural parameters used so far are accurate. In the following subsection, we show that this is not the case, and \rSVNO ~is a Mott insulator when the crystal structure parameters are calculated taking electronic correlations into account as well.

\begin{figure}
\centering
        \includegraphics[width=1.0\linewidth]{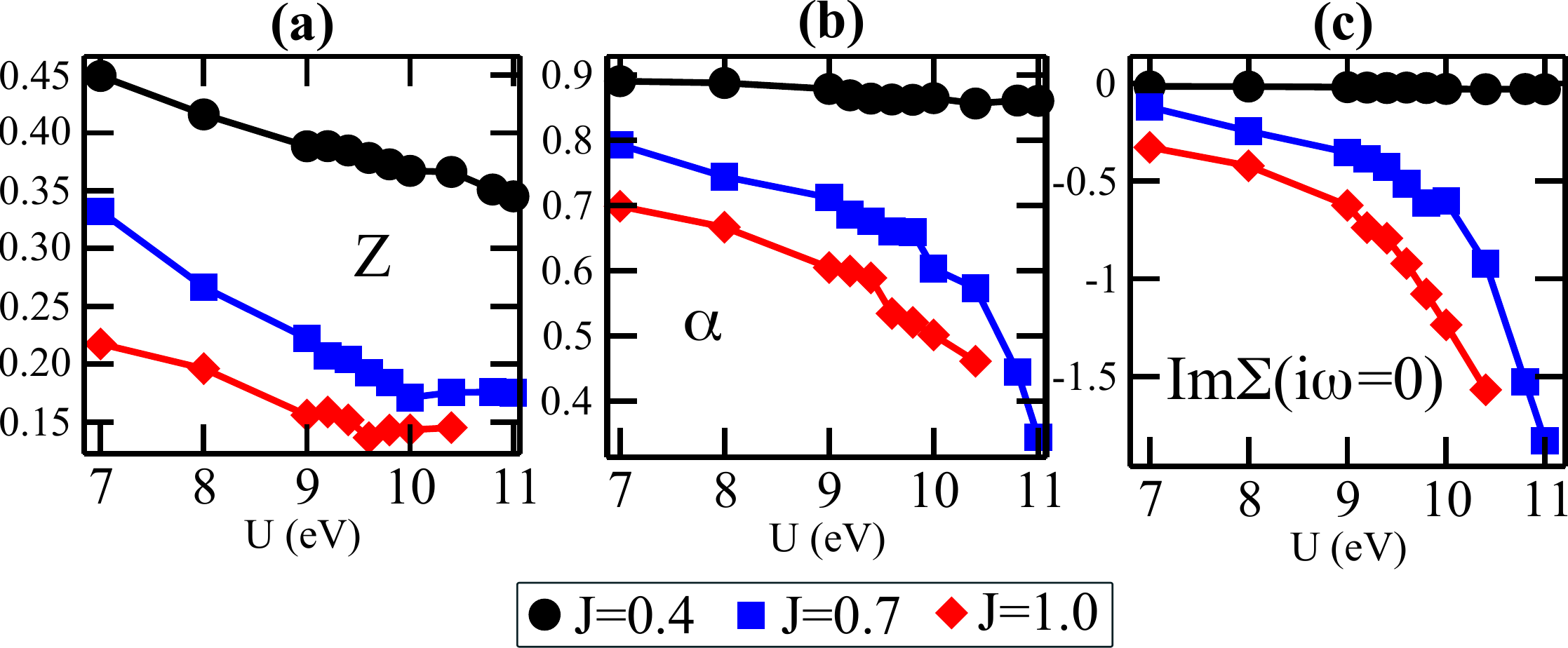}
		\caption{Dependence of (a) the mass renormalization ($Z$), (b) exponent of the real part of self energy ($\alpha$), and (c) zero energy intercept of the imaginary part of the self energy for V-t$_{2g}$ orbitals in \rSVNO ~as a function of $U$ for different Hund's coupling $J$ values from DFT+DMFT, calculated in the crystal structure as determined from DFT+U. For $J=1$~eV, the system becomes insulating above $U=10.4$~eV. }
        \label{fig:alpha_Z_im_U}%
\end{figure}%

To further elucidate the nature of the electronic correlations in metallic \rSVNO, we present the evolution of mass renormalization $Z$, self energy exponent $\alpha$, and the intercept of the imaginary self energy Im($\Sigma(i\omega=0$)) for the V-t$_{2g}$ orbitals in Fig. \ref{fig:alpha_Z_im_U} with Hubbard $U$ for three different values of $J$. While we classify this system as a Hund's metal, it is very close to the boundary of the Mott insulating phase, and increasing the value of Hubbard $U$ by 2~eV while keeping $J=0.7$~eV drives the system insulating. The critical $U$ value is mildly reduced by increasing $J$ to $J=1.0$~eV. Reducing the Hund's coupling to $J=0.4$~eV, on the other hand, has a more dramatic effect: it not only increases $Z$ by more than a factor of 2, but it also turns the system into a good Fermi liquid with zero intercept of the imaginary part of the self energy and exponent $\alpha$.

\subsection{Crystal structure of \rSVNO ~from DMFT} 

\begin{figure}
\centering
	\includegraphics[width=1.0\linewidth]{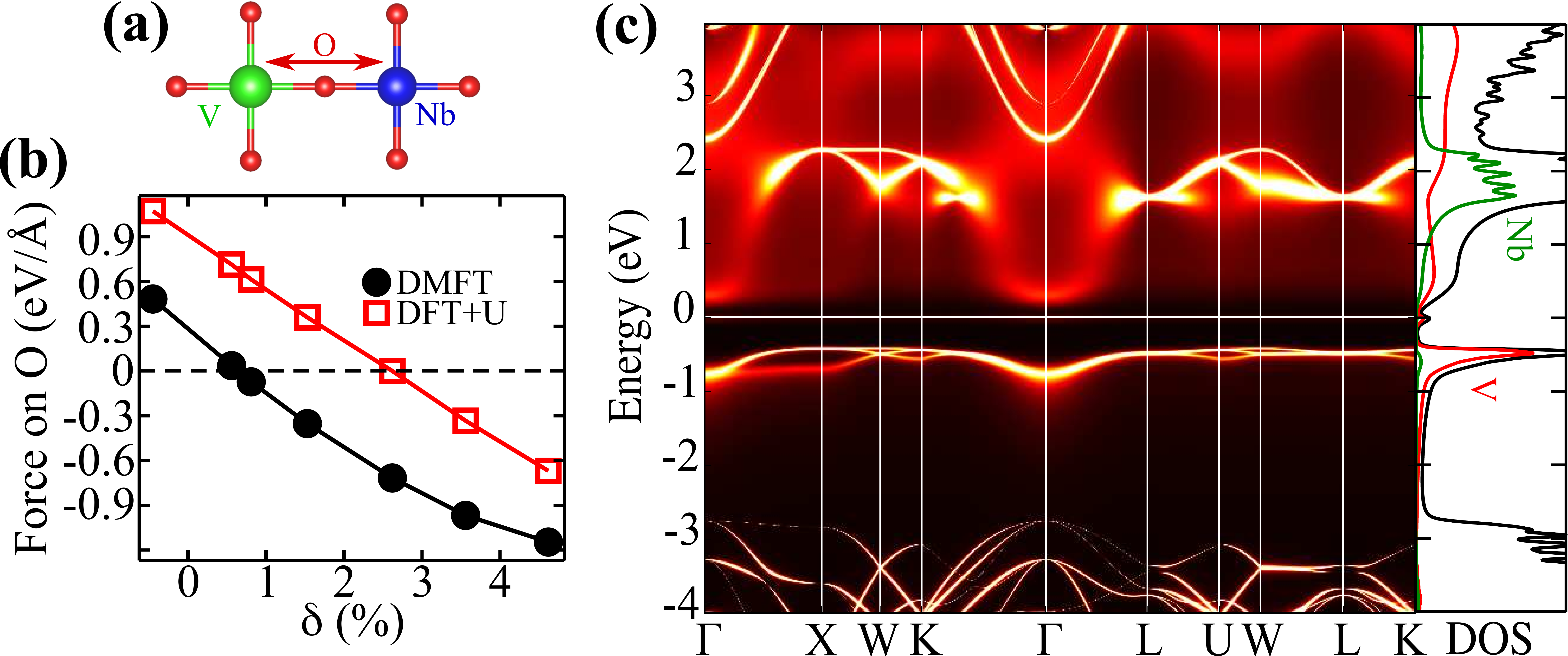}
		\caption{(a) The only internal parameter in the crystal structure of \rSVNO, the bond-length disproportionation. (b) The force on the O ions as a function of bond-length disproportionation $\delta$ calculated using DFT+DMFT, as well as DFT+$U$. The inset sketches how the only internal parameter $\delta$ changes the crystal structure. (c) Spectral function of \rSVNO ~from DFT+DMFT with $\delta$ also optimized using DFT+DMFT. }
        \label{fig:rs_forces}%
\end{figure}

The common approach in DFT+DMFT studies is to use the crystal structure obtained either from experiments or from DFT+$U$ calculations -- as we did so far. It has recently became also possible to obtain crystal structural parameters from DFT+DMFT using a stationary implementation \cite{Haule2015_energies, Haule2016Forces, Haule2018}. In order to check if the DFT+$U$ structural parameters are reliable in the room temperature paramagnetic phase, we optimized the ionic positions in rocksalt ordered \SVNO~using DFT+DMFT as well. There is only one internal parameter to optimize, which is the V--O/Nb--O bond length disproportionation, shown in Fig. \ref{fig:rs_forces}a. We quantify this disproportionation as $\delta=(d_{Nb-O}-d_{V-O})/(d_{V-O})$. In Fig. \ref{fig:rs_forces}b, we plot the force on the O atoms as a function of $\delta$ both from DFT+DMFT and DFT+$U$ for comparison. DFT+DMFT predicts a much smaller $\delta$ than DFT+$U$, which is in-line with the very similar ionic radii of V$^{3+}$ and Nb$^{5+}$ \cite{Shannon1976}. This 2\% reduction in $\delta$ has a strong effect on the electronic structure. The DFT+DMFT spectral function of \rSVNO~with $\delta=0.6\%$, shown in Fig. \ref{fig:rs_forces}c, indicates that this material is a Mott insulator. In other words, the effect of the electronic correlations on the crystal structure is strong enough to induce a transition from a Hund's metallic phase to a Mott insulating one in \rSVNO. The probabilities of V atomic impurity states (Supplementary Fig. S7 \cite{Supplement}) show that this transition is accompanied by an increase in $\langle |S_z|\rangle$ for vanadium, which also emphasizes the importance of Hund's coupling.  

In passing, we note that \rSVNO ~has Vanadium $S=1$ moments on a face centered cubic lattice. This is similar to other double perovskites La$_2$LiReO$_6$ and Ba$_2$YReO$_6$ where implications of frustrated magnetism were discussed \citep{Aharen2010, Vasala2015}. However, our DFT+DMFT calculations show no tendency for magnetic ordering at room temperature \cite{Supplement}, which is expected due to  the large distance and the lack of short superexchange pathways between the magnetic atoms in \rSVNO.  

\subsection{Layered ordered structure} 
While the rocksalt is the most common ordering type for B-site ordered double perovskites, another possibility is layered ordering \cite{King2010} (Fig. \ref{fig:str_dos}b). In bulk double perovskites, layered ordering is often stabilized by Jahn-Teller effect, which is strong in cuprates such as La$_2$CuSnO$_6$ and La$_2$CuZrO$_6$ \cite{Samanta2017, Azuma1998, Anderson1993}. While V$^{3+}$ has two electrons in its \t2g orbitals, and hence it can in principle lower its energy through a Jahn-Teller type distortion, this tendency is expected to be much weaker than that in Cu$^{2+}$ due to the difference in orbital characters. As a result, there is probably no driving force for layered ordering in bulk \SVNO. However, even for combinations of cations that form disordered solid solutions in bulk, layer-by-layer synthesis methods such molecular beam epitaxy make it possible to grow (001) heterostructures with periods as short as 2 unit cells, which can be considered as layered double perovskites. (For example Ref. \cite{Santos2011}.) The layered structure allows very different electronic properties since it hosts infinite 2 dimensional planes of either perovskite parent, which was the reason that the cuprate double perovskites were studied in a high T$_C$ superconductivity context \cite{Azuma1998, Anderson1993}. We now turn our attention to the layered ordered compound, which we refer to as \lSVNO. 

\begin{figure}
\centering
        \includegraphics[width=1.0\linewidth]{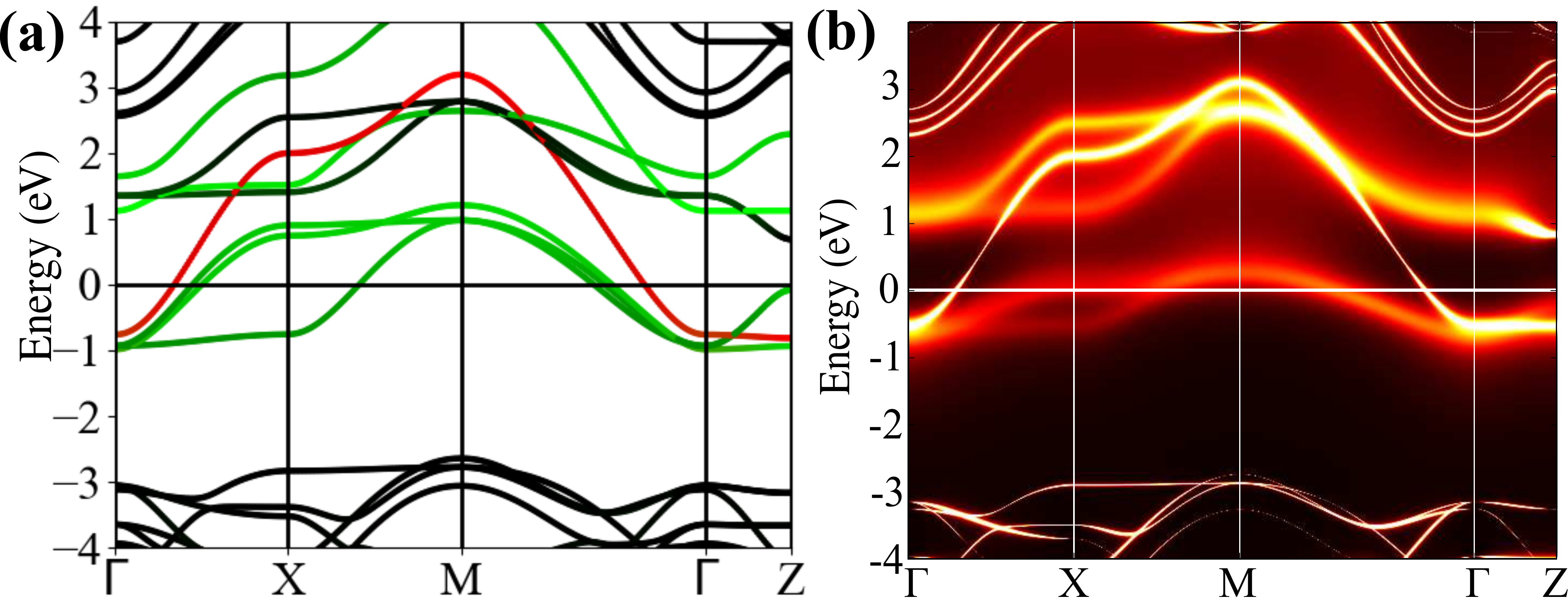}
		\caption{(a) DFT and (b) DFT+DMFT electronic structure of \lSVNO. In the DFT bandstructure, green and red colors represent total V and Nb-$d_{xy}$ character of the bands. }
        \label{fig:layered}%
\end{figure}%

In Fig. \ref{fig:layered}a, we show the DFT bandstructure of \lSVNO. Unlike \rSVNO, where only V bands cross the Fermi level, \lSVNO ~has some Nb states at Fermi level in addition to the V states. This is due to the large bandwidth of the Nb $d_{xy}$ band (where the $z$ axis is defined to be in the layering direction), which is highly dispersive in the NbO$_6$ planes because of the unbroken NbO$_6$ octahedral network in these planes. Thus, the V to Nb charge transfer is not complete in \lSVNO. 
The differences in the DFT bandstructures of \lSVNO ~and \rSVNO ~get more emphasized when electronic correlations are taken into account via DFT+DMFT. In Fig. \ref{fig:layered}b, we show the spectral function of \lSVNO ~obtained from DFT+DMFT, using internal ionic positions also predicted by DFT+DMFT. (The electronic correlations have a smaller effect on the crystal structure of \lSVNO ~than they do on that of \rSVNO ~\cite{Supplement}.) 
As a result of increased bandwidth and incompleteness of the charge transfer, \lSVNO ~doesn't become a Mott insulator when correlations are taken into account, as seen from the DMFT spectral function shown in Fig. \ref{fig:layered}b. 
This is a crucial difference between the two cation orderings: even though \rSVNO ~and \lSVNO ~consist of the same building blocks, the two different cation orderings result in different charge states and electronic structures according to both DFT and DFT+DMFT calculations. 
The width of the $d_{xy}$ band is renormalized by a small amount in \lSVNO, and its renormalization factor $Z_{Nb-xy}=0.86$ is close to that of bulk \SNO. The V-\t2g ~bands, on the other hand, exhibit stronger correlation effects with $Z_{V}\sim0.30$ for all three orbitals. The value of $Z$ depends on the Hund's coupling $J$ as expected, albeit weakly compared to \rSVNO: for the same range of $J$ values, $Z$ changes by less than two-fold, and for $J=0.7$~eV, the exponent of the imaginary part of the vanadium self energy is $\alpha_{V} \sim 0.84$, indicating mild Hund's metallic behavior \cite{Supplement}. 

Despite stronger correlations of the V--\t2g ~electrons than those of the Nb, \lSVNO ~is not a Mott insulator unlike the rocksalt ordered \rSVNO. This is due to the larger bandwidth of the transition metal bands crossing the fermi level, which is in turn due to the connected layers of VO$_6$ and NbO$_6$ layers in the layered compound. In other words, different cation ordering give rise to very different physical properties.
One interesting question that arises at this point is what the effect of cation disorder, i.e. random arrangement of V and Nb cations, is. First principles calculations that can simulate some degree of cation disorder require large supercells that are currently the beyond reach of DFT+DMFT due to their computational cost. Nevertheless, given that two different cations orders lead to metallic and insulating phases in \SVNO,  we posit that a metal-insulator transition should exist as a function of degree of cation ordering in \SVNO.

\section{Correlated metallic transparent conductors?} 

There is high demand for compounds that are transparent to visible light, and are good electrical conductors at the same time. Correlated metals can be promising as transparent conductors because of their large carrier concentration compared to doped semiconductors \cite{Zhang2016}. Both \SVO ~and \SNO ~have been considered as correlated transparent conductors with different transparency windows, where \SVO ~is transparent on the lower energy visible spectrum, and \SNO ~is transparent for blue and ultraviolet light \cite{Paul2019_PRM, Zhang2016, Park2020, SNO2}. A natural question to ask at this point is whether the layered \lSVNO ~double perovskite can bring together the positive aspects of both parent compounds and serve as a transparent metal that is transparent throughout the visible spectrum. Unfortunately, the optical conductivity (shown in the supplement \cite{Supplement}) of \lSVNO ~shows that this is not the case, because a wide absorption peak at $\sim 1.1$~eV emerges in the layered compound, hampering its use as a transparent conductor. However, a disordered or partially ordered \SVNO ~is likely to not have this absorption peak because of the lack of translational symmetry breaking, and might be more promising as a transparent conductor.

\section{Summary and conclusions} 

We performed first principles DFT+DMFT calculations on double perovskites of \SVNO ~with different (rocksalt and layered) cation orderings. Even though V and Nb are both group 5 transition metals, there is almost complete charge transfer between them when they coexist in the double perovskite. The two different double perovskites obtained by different orderings of V and Nb cations have starkly different electronic properties: The layered compound is a correlated metal with a mild or moderate degree of Hund's metallicity, whereas the rocksalt ordered compound is a Mott insulator with $S=1$ moments on a frustrated FCC lattice.

Our results underline two points that can be generalized to double perovskites beyond \SVNO: \textit{(i)} Inter-cationic charge transfer, which is quite common in double perovskites, can exist even between cations from the same column of the periodic table, and can give rise to very different chemistries than the parent perovskites.  \textit{(ii)} The strength of electronic correlations in double perovskites, as measured by $Z$ in our approach, can be tuned by the cation order. This might be a fruitful way to obtain new correlated metallic perovskite compounds. Finally, our results on \SVNO ~adds Hund's metallicity to the list of phenomena that emerges in double perovskites even when it does not exist in parent compounds.

\begin{acknowledgments}
This work is supported by NSF DMREF Grant DMR-1629260. We acknowledge the Minnesota Supercomputing Institute (MSI) at the University of Minnesota for providing resources that contributed to the research results reported within this paper.
\end{acknowledgments}

\end{document}